\documentstyle[12pt,epsf,draft]{article}

\newcommand\putfig[3]{
   \vbox{
   \let\picnaturalsize=N
   \def\picsize{#3}
   \def\picfilename{#1}
   \ifx\nopictures Y\else{\ifx\epsfloaded Y\else\input epsf \fi
   \let\epsfloaded=Y
   \centerline{\ifx\picnaturalsize N\epsfxsize \picsize\fi
   \epsfbox{\picfilename}}}\fi
   \vspace{1.0cm}
   {\it #2}
   \vspace{1.5cm}
   }
}

\def\be{\begin{equation}}
\def\ee{\end{equation}}
\def\bear{\begin{eqnarray}}
\def\eear{\end{eqnarray}}
\def\nn{\nonumber}

\def\ket{{\rangle}}
\def\hlf{{{1\over 2}}}

\def\wdg{{\wedge}}                              % wedge product

\newcommand\inv[1]{{1\over{#1}}}

\newcommand\rep[1]{{\bf {#1}}}      % representation
\newcommand\tr[1]{{\mbox{tr}\{{#1}\}}}          % trace
\newcommand\trr[2]{{\mbox{tr}_{#1}\{{#2}\}}}    % trace in a rep
\def\a{{\alpha}}

\def\u{{\mu}}
\def\v{{\nu}}

\def\lam{{\lambda}}

\def\Id{{\bf I}}                                % the Identity

\def\IZ{{\bf Z}}                                % Math Z
\def\dirac{{\Gamma}}                            % Dirac matrices in 6D

\def\TDL{{{\cal T}-duality\ }}                  % T-duality
\def\IIA{{IIA\ }}                               % Type IIA
\def\IIB{{IIB\ }}                               % Type IIB
\newcommand\MR[1]{{{\bf R}^{#1}}}               % Real numbers
\newcommand\MS[1]{{{\bf S}^{#1}}}               % Circle, sphere,...
\def\wdg{{\wedge}}                              % ^
\def\ExE{{$E_8\!\otimes\! E_8$}}                % E8 times E8
\def\Mth{{M-theory}}                            % M- theory
\def\npb#1#2#3{{\it Nucl.\ Phys.} {\bf B#1} (19#2) #3}
\def\plb#1#2#3{{\it Phys.\ Lett.} {\bf B#1} (19#2) #3}

\def\prd#1#2#3{{\it Phys.\ Rev.} {\bf D#1} (19#2) #3}

\def\hepth#1{{\it hep-th/{#1}}}

\begin{document}
\begin{titlepage}
\titlepage
\rightline{IASSNS-HEP-96/12}
\rightline{PUPT-1595}
\rightline{hep-th/9602120}
\rightline{Feb 21, 1996}
\vskip 1cm
\centerline {{\LARGE  Small $E_8$ Instantons}}
\centerline {{\LARGE  and Tensionless Non Critical Strings}}
\vskip 1cm

\centerline {Ori J. Ganor}
\vskip 0.5cm
\begin{center}
\em  origa@puhep1.princeton.edu\\
Department of Physics, Jadwin Hall, Princeton University\\
Princeton, NJ 08544, U. S. A.
\end{center}
\vskip 1cm
\centerline {and}
\vskip 0.5cm
\centerline {Amihay Hanany}
\vskip 0.5cm
\begin{center}
\em hanany@sns.ias.edu\\
School of Natural Sciences, Institute for Advanced Study, 
Olden Lane,
Princeton, NJ 08540, U. S. A.
\end{center}
\vskip 1cm
\abstract{
T-duality is used to extract information on an instanton
of zero size in the $E_8\times E_8$ heterotic string.
We discuss the possibility of the appearance of a tensionless anti-self-dual
non-critical string through an implementation of the mechanism
suggested by Strominger of two coincident 5-branes.
It is argued that when an
instanton shrinks to zero size a tensionless non-critical string
appears at the core of the instanton.
It is further
conjectured that appearance of tensionless strings in the spectrum leads
to new phase transitions in six dimensions in much the same way as massless
particles do in four dimensions.
}
\end{titlepage}

%%%%%%%%%%%%%%%%%%%%%%%%%%%%%%%%%%%%%%%%%%%%%%%%%%%%%%%%%%%%%%%%%%%%%%%%%%%%
%                      PAPER                                               %
%%%%%%%%%%%%%%%%%%%%%%%%%%%%%%%%%%%%%%%%%%%%%%%%%%%%%%%%%%%%%%%%%%%%%%%%%%%%
\section{Introduction}

Recently Witten \cite{WitSML} showed what happens when
instantons on the $SO(32)$ heterotic string shrink to zero size.
The low energy six dimensional effective theory has
an extra $Sp(1)$ gauge symmetry which is supported at the core of
the instanton. The finite size instanton is obtained from the zero
size instanton through a Higgs mechanism.
In six dimensions there is no vector multiplet moduli space and
therefore
the Higgs mechanism is presumed to be the only relevant low energy
dynamics.

In this paper we will explore some features in the problem of small
instantons on the \ExE\ heterotic strings.
A straight forward generalization of Witten's construction does not
seem to work because the dimensions of the $E_8$ representations are
too big.
It was suggested by Witten \cite{WitLEC} that the solution involves a
new type of physics unknown before.

In \cite{DLncs} Duff and Lu constructed a supersymmetric solitonic
one-brane solution of $N=2$ $D=6$ supergravity which is self-dual.
This soliton was realized in \cite{WitCOM} as a self-dual three brane
of the type IIB string compactified on K3 wrapped on a self-dual
two-cycle of the K3. When the K3 degenerates in such a way that the
pullback of the complexified K\"ahler form on the two cycle is very
small the tension of the one-brane (which is proportional to the area
of the two cycle) is much smaller than any other scale in the theory.
Another realization of this one-brane was discussed in \cite{StrOPN}.
A membrane stretched between two five-branes in the \Mth\ has the
one-brane solution at the boundaries of the membrane. When the two
five-branes coincide the tension of the one-brane vanishes.
The constructions of \cite{StrOPN} and of \cite{WitCOM} are related as
explained in \cite{WitFIV}.

Returning to low energy in six dimensions we see that
a new type of dynamics may appear through
the emergence of a tensionless self-dual string.

We asked ourselves the following questions:
\begin{itemize}
\item
What can we learn about the \ExE\ small instanton
from T-duality between $SO(32)$ and \ExE?
\item
Is there any relation between tensionless strings in 6D
and an \ExE\ small instanton?
\item
What happens when a 5-brane of the \Mth\ on $\MS{1}/\IZ_2$
(which describes the strong coupling limit of the \ExE\ heterotic
string)
approaches one of the fixed points?
\end{itemize}

The paper is organized as follows.
Section (2) is a review of small instantons in the $SO(32)$ heterotic
string. In section (3) we employ T-duality to argue the existence of a
tensionless string in the low energy limit for a small \ExE\ instanton.
In section (4) we derive a geometrical picture for the tensionless
string from the \Mth.
Section (5) discusses the relation to D-branes in type I and type IA.
In the appendix we calculate the anomalies which are needed for the
geometrical description of section (4).

\section{Review of small instantons in the $SO(32)$ heterotic string}
One of the ways in which string perturbation theory breaks
down is by approaching singularities of the gauge bundle.
This is what happens when the heterotic string is compactified 
down to 6D in such a way that the gauge bundle on the four compactified
dimensions has an instanton whose size is such that the curvature at
its core cannot be neglected.
The limit of taking the size of the instanton to zero produces
a soliton with a region in which the dilaton blows up.
Those solutions were constructed in 
\cite{StrHSO,CHS1,CHS2,CHSREV}.

In \cite{WitSML} the behavior of small instantons in the $SO(32)$
heterotic string was described. It was found that when a small
instanton shrinks to zero size an extra gauge symmetry appears that is
supported at the core of the instanton. In addition, hypermultiplets
which transform in the $(\rep{2},\rep{32})$ representation of
$Sp(1)\otimes SO(32)$
as well as a singlet field appear in the massless spectrum as the
instanton shrinks. When $k$ instantons shrink at the same point the
gauge group is $Sp(k)$ and the massless hypermultiplets transform in
the representations $(\rep{2k},\rep{32})$ of $Sp(k)\otimes SO(32)$ and
the antisymmetric of
$Sp(k)$ (which is reducible and decomposes to a singlet plus the 
irreducible $\rep{(2k^2-k-1)}$).

\section{Deductions from T-duality}

The subject of our discussion
is the zero size instanton in the heterotic $E_8\times E_8$ string,
but in order to be concrete let us think of it as compactified on K3,
as in \cite{WitSML}.
The coordinates $x_1,\dots x_6$ will be the uncompactified
dimensions on $\MR{6}$, $x_1$ denotes the time coordinate.
The six dimensional theory has a $(2,0)$ supersymmetry which
is the minimum possible in 6D. (There are two left spinor generators
and no right spinor generators. This reduces to $N=2$ in 4 dimensions
which is really a $(2,2)$ supersymmetry.)
Among other data the theory is defined with a left moving
gauge bundle with instanton number 24. We will be interested in
singularities in moduli space for which one or more instantons shrink
to zero size.
We assume that the gauge fields of the instanton
are embedded in one of the two $E_8$-s.
Taking the limit where the size of
K3 is very large we can, as in \cite{WitSML}, forget about
the K3. The instanton can be thought of as being on $\MR{4}$.

Ten-dimensional \ExE\ and $SO(32)$ heterotic strings are
distinct, but when compactified on a circle $\MS{1}$ of finite
radius their moduli spaces are identical and both heterotic
theories are in fact equivalent under \TDL\ \cite{GinTOR}.

It is therefore natural to ask how the \ExE\ heterotic theory
compactified on K3, with an instanton that has shrunk, behaves
upon further compactification on $\MS{1}$ down to 5D.

In what follows we would like to argue on the basis of T-duality
that relates the $SO(32)$ heterotic string to \ExE\ heterotic string,
that there exists a string-like object with zero-tension
in the six-dimensional theory (that is the low-energy theory
of an observer in the uncompactified dimensions).
We recall that such a string-like object first appeared in
type-\IIB compactified on a K3 with a shrinking 2-cycle \cite{WitCOM}
and was called the ``non-critical string''~\cite{DLncs}.

\subsection{T-duality facts}
Starting with an \ExE\ string on $\MR{9}\times\MS{1}$
where the $\MS{1}$ is of radius $r$, we add the special Wilson loop
on the $\MS{1}_r$ of the form (in the adjoint representation
$\rep{248}$ of $E_8$):
\be
W= \left(
\begin{array}{cc}
 \Id_{120\times 120}   &    0    \\
          0            &    -\Id_{128\times 128} \\
\end{array}
\right)
\label{WilE8}
\ee
This breaks \ExE\ down to $SO(16)\times SO(16)$.
According to \cite{GinTOR}, the result is equivalent to an $SO(32)$
heterotic string on $\MR{9}\times\MS{1}$ where $\MS{1}$ is of
radius $\inv{r}$ with the Wilson loop  (in the fundamental
representation $\rep{32}$ of $SO(32)$):
\be
W= \left(
\begin{array}{cc}
 \Id_{16\times 16}   &    0    \\
          0          &   -\Id_{16\times 16} \\
\end{array}
\right)
\label{WilSO32}
\ee
Which also breaks $SO(32)$ down to $SO(16)\times SO(16)$.
This is the same Wilson loop that was used in \cite{PolWit,HorWit}.
\subsection{Application to small instantons}
When we compactify one direction of $\MR{6}$ (in $\MR{6}\times K3$)
so that the six-dimensional small instanton wraps around the $\MS{1}$
of $\MR{5}\times\MS{1}\times K3$, massless states of the
six-dimensional
theory will become BPS saturated states of the five-dimensional theory
on $\MR{5}$. $N=2$ supersymmetry in 5D has one {\em real} central
charge which is invariant under T-duality. 
The special Wilson loops $W$ that were chosen above have the
virtue of mapping  the $SO(16)\times SO(16)$ symmetry
group on one  side (say heterotic on \ExE) to the group on the
other side (heterotic on $SO(32)$). So $SO(16)\times SO(16)$
quantum numbers of BPS states will not change under T-duality.

This is not the only good quality of $W$, and we will soon
make use of its other properties as well.

We start with some massless state that exists in the
``mysterious'' six-dimensional theory of the \ExE\ small instanton.
Compactifying on a radius $r\gg 1$ (with the special
Wilson loop $W$) we expect to get a BPS
state in the 5-dimensional theory which will become a BPS
state with {\em the same} $SO(16)\times SO(16)$ quantum numbers
in the $SO(32)$ theory compactified on a radius $\inv{r}$.
Conversely, if we knew what are the BPS states of the $SO(32)$ theory
compactified to 5-dimensions on a radius $r' \ll 1$ we would
expect to find massless states with the same $SO(16)\times SO(16)$
quantum numbers in the mysterious theory.

We know from \cite{WitSML} that on $\MR{6}$ the $SO(32)$ zero-size
instanton has an additional $Sp(1)$ gauge symmetry
and massless hyper-multiplets in the $(\rep{32},\rep{2})$ 
representation of $SO(32)\times Sp(1)$. Thus, for $r'\gg 1$
the same theory on $\MR{5}\times \MS{1}_{r'}$ has BPS
states with the quantum numbers $(\rep{16},\rep{2})$ (with
the $\rep{16}$ corresponding to the $SO(16)$ in which the instanton
was embedded).
Note that from the explicit form (\ref{WilSO32}) of $W$ it
follows that the other $(\rep{16},\rep{2})$ fields which
are in the fundamental representation of $SO(16)$ are multiplied by
$-\Id_{16\times 16}$ in (\ref{WilSO32}). They thus have
anti-periodic boundary conditions along the $\MS{1}$ and
they give rise to massive states in 5D, which become very heavy
once we decrease the radius of $\MS{1}$.
\footnote{We are grateful to E. Witten for pointing this out
as well as the special features of $W$ in the next paragraph.}

What happens as we decrease $r'$?
We know from \cite{CCDF,AFT} that in the general case
the vector-multiplet
moduli space of the heterotic string on $K3\times \MS{1}_{r'}$
has a ``jump'' singularity at $r'=1$.
BPS states can become unstable even at smooth points
of the vector-multiplet moduli space (as in \cite{SeiWit}).
However, there will always be at least one BPS state (probably
even with zero mass) in the $(\rep{16},\rep{2})$ 
which has nothing to decay into. It will be stable.
Actually, the special Wilson loop $W$ that we used makes life easier.
According to \cite{PolWit} the only Wilson line for which there is no
extra
enhanced gauge symmetry at some value of the radius is the Wilson line
given by equation (\ref{WilSO32}). We therefore expect no transitions as
we take the limit of small radius.

We are led to the conclusion that the \ExE\ small instanton
compactified on $\MS{1}$ contains massless states in the 
representation $\rep{16}$ of $SO(16)$.
In six-dimensions, before the compactification on $\MS{1}$,
we expect
$E_8$ to be a gauge symmetry of the theory of the zero-size
instanton. However, when we decompose the representations
of $E_8$ under the subgroup $SO(16)\subset E_8$, the fundamental
representation $\rep{16}$ of $SO(16)$ never appears!

Thus, the $\rep{16}$ states could not have been there
{\em before the compactification}.

We find that there are massless states which appear
when the theory is compactified on $\MS{1}$ with an arbitrarily
big radius, but there are no corresponding massless states
when the theory is not compactified.

Such a situation was, of course, encountered before.
We remember it from type-\IIB compactified on K3, discussed
in \cite{WitCOM}. This was the first appearance of the 
{\em tensionless non-critical string}. It was explained in \cite{WitCOM}
that a string with zero tension will produce such an effect.
It is a question of ``the order of the limits''.
A winding state of a string of tension $\epsilon$ on a circle
of radius $R$ will produce a state of mass of order $\epsilon R$.
When we take $R\rightarrow\infty$ first, the state acquires infinite
mass. When we take $\epsilon\rightarrow 0$ first, the state becomes
massless.
 
In the next section we will see where this tensionless
string ``comes from'' and explore what happens when
we lift the tension $\epsilon$.

\section{Strong coupling limit}

Since the K3 is very large, we can replace
it locally by $\MR{4}$
with a solitonic 5-brane charged under the dual of the (NS-NS field)
$B_{\u\v}$ sitting at the origin of the $\MR{4}$
(and having all values of $x_1,\dots x_6$).
Taking, in this situation, the asymptotic value of $\lam_{st}$
(i.e. at infinity of $\MR{4}$) to infinity doesn't change
the nonperturbative phenomena that originate ``from deep down
the throat''. 
This brings us to the \Mth\ compactified on the orbifold
$\MS{1}/\IZ_2$ with the radius of $\MS{1}$ becoming large
\cite{HorWit}.
The charge of the soliton corresponds
to the charge of the ``elementary'' 5-brane of the \Mth\ \cite{WitNEW}.
We will assume that the small instanton becomes a 5-brane of the \Mth.
According to \cite{HorWit}
the gauge fields of the \Mth\ ``live''
only on the fixed points of $\MS{1}/\IZ_2$ in such a way that
each $E_8$ in \ExE\ comes from a different fixed point.
Thus, in order for the 5-brane to describe massless states
that are charged under one of the two $E_8$-s
 it must lie entirely on one fixed point.
Since the distance between the two fixed points
increases with the 10D coupling constant, and we are taking it to
infinity, we end up with the \Mth\ on $\MR{6}\times\MR{4}\times\MR{+}$
with a 5-brane that sticks to the boundary at the position 
$\MR{6}\times {\vec{0}}\times {0}$ (the first $\vec{0}$ is the origin
of $\MR{4}$ and the second is the fixed point of
$\MR{+}=\MR{1}/\IZ_2$.)

How is the $(2,0)$ SUSY structure obtained?
The 5-brane in 11D has $(4,0)$ SUSY on its world-volume
(Indeed, it supports a multiplet comprising of a anti-self-dual two-form
and 5 scalars). The extra $\IZ_2$ imposed by the orbifold in the $x^{11}$
direction leaves $(2,0)$ for the uncompactified six dimensions.

\vskip 0.5cm
\putfig{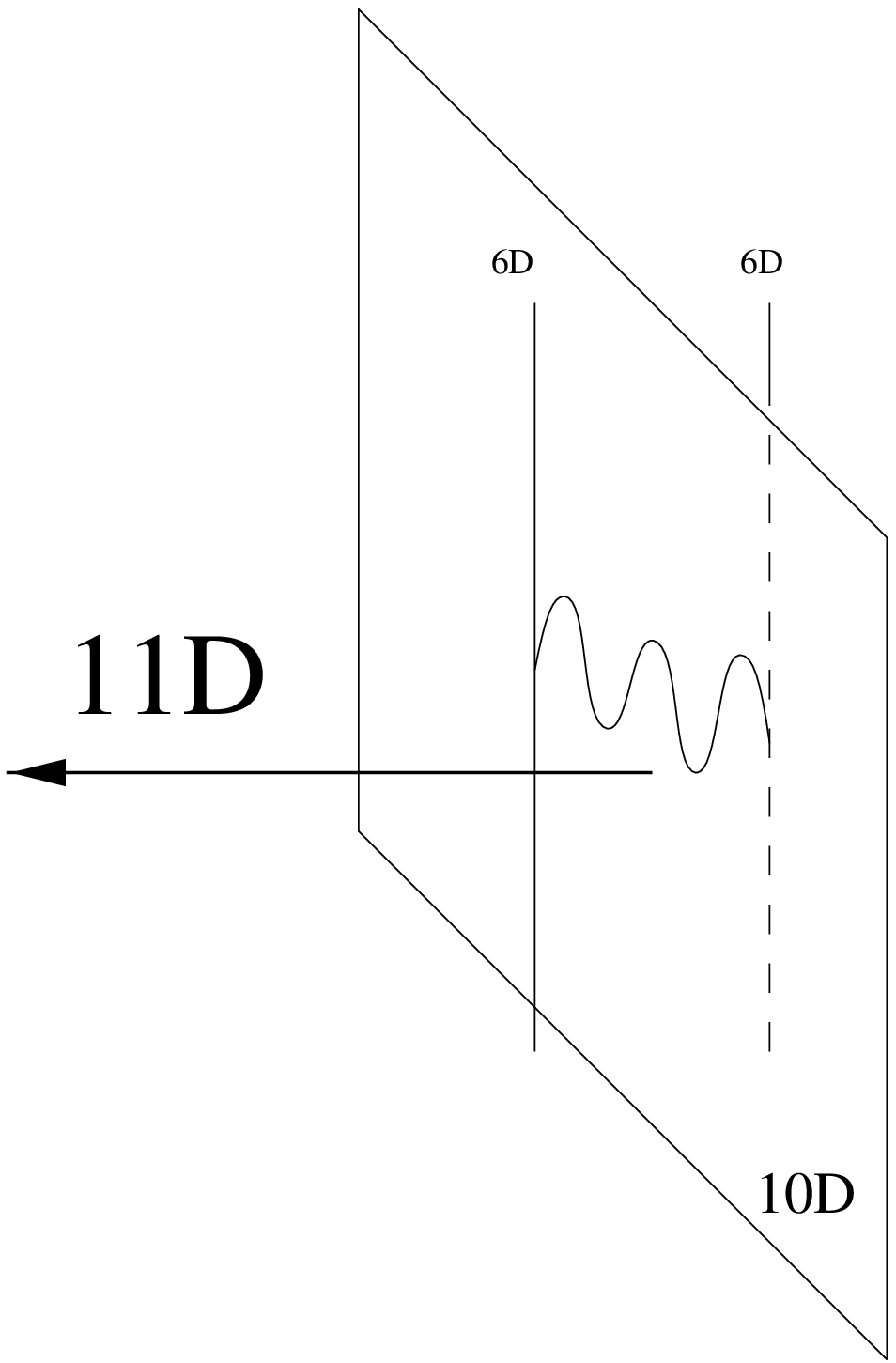}{\it Fig. 1:
        A 5-brane close to the boundary of 11D and its image,
        with a 2-brane connecting them.}{60mm}
\vskip 0.5cm

\subsection{Appearance of a tensionless string}
The setting is almost the same as in \cite{StrOPN}.
There we learn that a membrane can end on five-branes (see also the discussion
in \cite{Becker}).
 From \cite{HorWit} we also know that the membrane can end on one of 
the $\IZ_2$ fixed points (which might be called a 9-brane).

Let us briefly review the construction of \cite{StrOPN}.
Consider a membrane stretched between two five-branes in
eleven dimensional supergravity.
The low energy dynamics of the five brane is described by an $N=2$ $d=6$
chiral tensor multiplet containing $5$ scalars and an anti-self-dual
antisymmetric tensor. We denote its field strength by $T$.
The electric charge of the membrane with respect to the $3$-form potential
is given by integrating over an $S^7$ which surrounds the membrane
$$Q=\int_{S^7}*F,$$
$F$ being the $4$-form field strength.
The boundary of the membrane is a string lying inside the five-brane
world-volume.
This string serves as a source for the anti-self-dual antisymmetric tensor
$$Q=-\int_{S_3}T.$$
When the five-branes are separated the low energy dynamics is given by
2 tensor multiplets and their moduli space is given locally by a symmetric
space $SO(5,2)/(SO(5)\times SO(2))$.
When the positions of the five branes coincide a tensionless string
arises which carries no charge with respect to the $SO(2)$ group that
acts on the two five-branes.

What happens in our case?
We wish to identify the zero size instanton in the heterotic theory
with  an \Mth\ configuration which is the limit of a 5-brane
approaching the boundary (the $\IZ_2$-fixed point which we will later
also refer to as the 9-brane) until it becomes submerged in it.
Implementing what we studied in \cite{StrOPN} and \cite{HorWit} we
deduce that two kinds of ``new'' massless excitations can appear.
Those are a 2-brane stretched between the 5-brane and its $\IZ_2$
mirror image and a 2-brane that starts on the 5-brane and end on the
9-brane. In the limit when the distance from the 5-brane to the 9-brane
is zero, both excitations will be observed as
 1-branes of zero tension in the uncompactified 6D.

In the rest of the paper we will support the above argument
with some further tests:
\begin{itemize}
\item
The instanton tension scales like $1\over\lambda^2$ and the five-brane has
the same behavior (does not depend on the radius) after a Weyl rescaling.
\item
We will discuss the appearance of the hypermultiplet in the
$\rep{16}$ of $SO(16)$ after compactification on $\MS{1}$ and
breaking of \ExE\ down to $SO(16)\times SO(16)$.
\item
After the $\MS{1}$ compactification, it is possible
to transform to the type-I string as in \cite{HorWit}. 
We will identify the counterpart of the proposal in the type-I
language.
\item
Collapse of $k$ small instantons at the same point will be discussed as well
as its relation to type I and small $SO(32)$ instantons.
\item
Separation of the 5-brane from the 9-brane will be identified
as an $Sp(1)$ Wilson loop in the T-dual $SO(32)$ theory
(again after further compactification).

\end{itemize}

\subsection{low-energy fields}
In the previous section we described the low energy spectrum of a membrane
stretched between two five-branes ($5-5$ membranes). We now turn to describe
the low energy fields of $5-9$ membranes.
So we consider a membrane stretched between a five-brane and a nine-brane as in
figure $2$. Note that a small instanton differs from
this configuration by having the five-brane located at the nine-brane.
We nevertheless, for the propose of describing the low energy fields, keep
the branes apart.

\vskip 0.5cm
\putfig{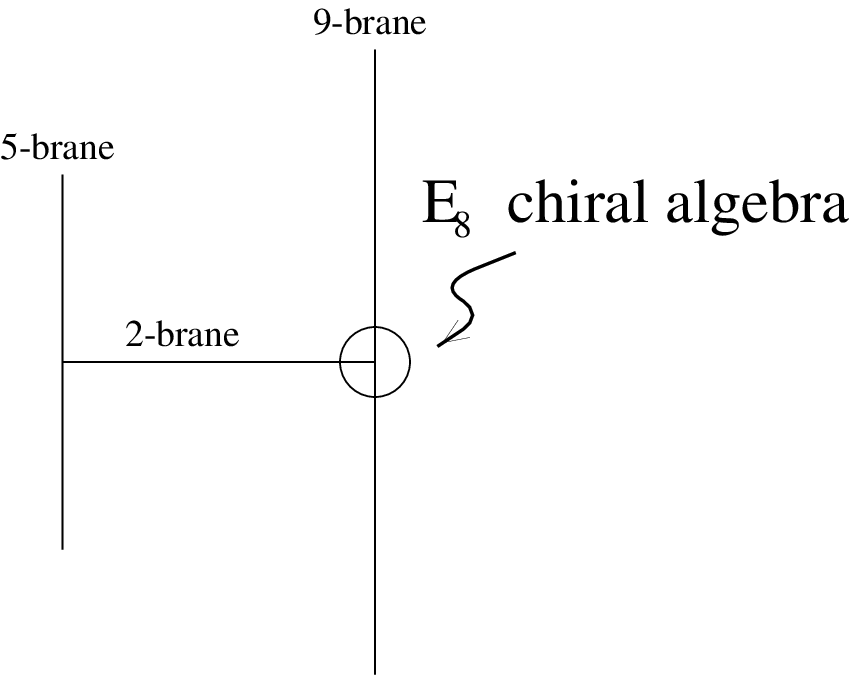}{\it Fig.2:
        A chiral $E_8$ that appears on the intersection
        of a 2-brane with a fixed point}{60mm}
\vskip 0.5cm

%\subsection{The low-energy intersecting field-theory}
Let us analyze the configuration of Fig.2, where
we have the \Mth\ on $\MR{1}/\IZ_2$ (that is
$\MS{1}/\IZ_2$ near the $x^{11}=0$ fixed point, with the 
other fixed point very far away so that we can forget about it).
We also have a (cosmic) 5-brane at distance $r$ from the fixed point
(i.e. at $x^{11} = r$ and say $x^7 = \cdots = x^{10} = 0$).
We assume that $r\gg {\a'}^{1/2}$ and that there is a cosmic 2-brane
stretched between the 5-brane and the fixed point, that is at
position
\be
x^3 = x^4 = \cdots = x^{10} = 0,\qquad
0\le x^{11} \le r
\ee

It is now legitimate to ask how an eleven-dimensional observer at a scale
that is much larger than ${\a'}^5$ describes the low-energy physics.
The low-energy field theory comprises of various fields.
First there are the 11D supergravity fields  that live in the 11D bulk.
Then there are 10D fields that live just on the ten dimensional
hyper-surface $x^{11}=0$ and interact with the boundary value
of the 11D fields on $x^{11}=0$. 

The physics at the $x^{11}=0$ end
was worked out in \cite{HorWit} on the basis of anomaly cancelations
both in 11D and on the 2-brane. It was argued there that
on the fixed point there is an $E_8$ 10D Super Yang-Mills theory,
and on the 2D intersection there is a chiral level one $E_8$ 
current algebra.
On the 2-brane bulk there lives the dimensional reduction of 
super Yang-Mills from 10D down to 3D (see \cite{WitBST}).
In 3D we get 8 scalars (using
the fact that a vector is dual to a scalar in 3D) that
represent transverse oscillations.
Now we come to the question of what lives on the 5-brane.
Generically, there lives
a tensor multiplet and a hyper-multiplet --
together we get one antisymmetric tensor field $B_{mn}^{+}$ (with
an anti-self-dual field strength) and 5 scalars
that represent transverse oscillations of the 5-brane.
However, when the 5-brane is stuck on
the boundary (9-brane), the situation changes a little bit.
Out of the 5 scalars, the one that corresponds to a translation
in the $x^{11}$-th direction, i.e. away from the boundary,
is no longer a modulus.
When we decompose the $(4,0)$ SUSY multiplet of the $B_{\u\v}^{(+)}$
and 5 scalars under the surviving $(2,0)$ SUSY we find two multiplets.
One comprises of $B_{\u\v}^{(+)}$ and one scalar, and the other
is the usual hyper-multiplet of 4 scalars. Since the modulus of
the first one corresponds to $x^{11}$ translations and is frozen
as we argued above, we do not see the tensor multiplet in the
resulting 6D theory (see also the comment in \cite{WitNEW}).
The exact mechanism by which the tensor multiplet becomes
``frozen'' is not clear to us, but it must involve 
tensionless string dynamics, since as we will see below, after
compactification to 5D on an arbitrarily big radius the transverse
directions re-appear and correspond to $Sp(1)$ Wilson loops
along the 6th compactified direction.
In fact in section (5.2) we will conjecture that in six dimensions the
small instanton corresponds to a point in the moduli space that
connects two different phases. One phase is in which the instanton has
nonzero size. The other phase was discussed in \cite{DMW} where it was
pointed out that the anomaly can be cancelled by a gauge bundle with
instanton number 23 (i.e. one unit less) and a five brane located in an
arbitrary position in the $x^{11}$ direction.

The remaining question is what lives on the intersection of the
5-brane with the 2-brane. We show in the appendix that
anomaly cancelation requires an anomaly free 2D theory.
So to summarize we have the following fields
\begin{enumerate}
\item 
11D supergravity in the bulk
\item
Super $E_8$ Yang-Mills in 10D on the fixed point
\item
A tensor multiplet and a hyper-multiplet in 6D on the
5-brane bulk.
\item
Dimensional reduction of $U(1)$ super-YM down to 3D on the
2-brane bulk
\item
A chiral $E_8$ level one current algebra on the 2D boundary of 
the 2-brane on the fixed point.
\item
We assume that there is nothing on the boundary where the 2-brane
ends on the 5-brane.
\end{enumerate}

\subsection{Recovery of the fundamental representation}
We argued previously that the effective theory in 6D
is governed by a tensionless string, whereas compactification
on $\MR{5}\times\MS{1}$, where the $\MS{1}$ is arbitrarily large,
yields an effective {\em field} theory in 5D. When we compactified
on $\MS{1}$ with the special Wilson loop $W$ of (\ref{WilE8}), 
we obtained the theory of \cite{WitSML} and in particular
hypermultiplets in the $\rep{16}$ of $SO(16)$ appeared.
Since we know from the previous subsection what are the
various low-energy fields, let us try to recover
the $\rep{16}$ states. 
So we compactify the picture of the previous section on $\MS{1}$
with the special Wilson loop $W$.
In the previous picture the
5-brane was separated from the fixed-point.
We will argue 
in section (5) that in the $SO(32)$ picture of \cite{WitSML} the separation
corresponds to an $Sp(1)$ Wilson loop along $\MS{1}$. We will see there that
the $(\rep{2},\rep{16})$ states that we are looking for have a mass
which is proportional to the Wilson loop and to the separation
in the \Mth\ picture. They are still in a reduced representation,
and so are BPS states.
The BPS argument allows us to work with the separated 5-brane and 9-brane.
BPS excitations of D-branes correspond to the {\em vacuum} of
the low-energy that lives {\em on the D-brane}. Thus, we have
to calculate the vacuum degeneracy of the low-energy theory
of the previous subsection.

We realize the $E_8$ chiral algebra that lives on the 2D boundary
at the fixed point by 16 chiral fermions (with both NS and R
boundary conditions).   We recall that the $E_8$
ground states are
\be
\psi_{-\hlf}^i\psi_{-\hlf}^j |0\ket
\ee
(120 states in the NS sector), plus the 128 states 
$|\a\ket$ in the R sector. Because of a normal ordering constant
all the 248 states have zero energy.
After the Wilson loop only the 
$\psi_{-\hlf}^i|0\ket$ states are the surviving low-energy excitations.
This is because:
\begin{enumerate}
\item
The Wilson loop
adds a normal-ordering contribution to $L_0$ which
changes the $\psi_{-\hlf}^i|0\ket$ from tachyonic to massless.
\item
The GSO projection 
projected out the $\psi_{-\hlf}^i|0\ket$ and left the $|0\ket$
before the addition of the Wilson loop.
The operators $\psi^i$ are invariant under $W$, but since $W$ acts
as $(-1)$ on the states in the R sector, it means that the GSO charge
operator $(-)^F$ has an extra $(-)$ sign when acting on the vacuum
$|0\ket$. The result is that $\psi_{-\hlf}^i|0\ket$ are not projected
out while the tachyonic $|0\ket$ is projected out (see \cite{CJP}
for a related discussion).
\end{enumerate}
 The $\rep{2}$ in the 
$(\rep{2},\rep{16})$ just corresponds to wrapping the 2-brane in
opposite directions around the $\MS{1}$.

\section{Relation to type-I}
What is the physical meaning, for the heterotic string, to 
endowing the tensionless string with a small tension of $\epsilon > 0$?
We need to know what happens to the heterotic string when
we separate the 11D 5-brane from the fixed point (of $\MS{1}/\IZ_2$).
We begin by asking ourselves
how the Chan-Paton factors of two, discovered in
\cite{WitSML} manifest themselves in $E_8$ instantons in the M-theory.

To relate $E_8$ theory instantons to type-I Dirichlet 5-branes
we will follow exactly the same steps as in \cite{HorWit}.
First we start with \Mth\ on $\MS{1}/\IZ_2$. As in \cite{HorWit},
it must be mentioned that the $\IZ_2$ doesn't act {\em only} on 
the $\MS{1}$ but is a {\em combined} action of a reflection of 
the $\MS{1}$ and
\be
A_3 \longrightarrow -A_3
\ee
where $A_3$ is the 3-form potential of 11D supergravity.
This transformation means that the $\IZ_2$ reverses the
world-volume orientation of the fundamental 2-brane in 11D
(to which $A_3$ couples). The orientation of the 5-brane
is invariant because it couples to the dual to $A_3$
(which is invariant because of an 11D $\epsilon$-symbol in the
formula). The above transformation of $A_3$ is necessary 
in order to find a surviving 2-form in 10D which would couple
to the fundamental heterotic string \cite{HorWit}.

As explained in \cite{HorWit}, after compactification on another
$\MS{1}$ down to 9D, we find a theory that is an orbifold of type-\IIA
on $\MS{1}/\IZ_2$, but the $\IZ_2$ acts both on the $\MS{1}$
and exchanges the orientation of the type-\IIA worldsheet
(because the type-\IIA string is the wrapped 2-brane of the \Mth).
This theory was called the type-IA. T-duality (which inverts the radius
of the $\MS{1}/\IZ_2$) brings this theory to the usual type-I.
Now, insert a 5-brane in type-I. According to \cite{WitSML} this
5-brane has a Chan-Paton factor that can have two values.
There are world-volume
$Sp(1)$ gauge fields living on the type-I 5-brane. The 
D-brane description \cite{PolDBR} of the 5-brane assigns Neumann
boundary conditions in the $\MS{1}$ direction (on which
the type-I 5-brane is wrapped).
T-dualizing this configuration we go to the type-IA,
in which there are orientifold planes at each
fixed point $x^{11}=0,\pi$. The $x^{11}$ coordinate in type-IA
corresponds to an $SO(32)$ Wilson line eigenvalue, in the sense
that turning on an $SO(32)$ Wilson line in type-I will cause
a translation in the position of 
the 32 8-branes in the $x^{11}$ direction in type-IA.
The type-I 5-brane which was wrapped on the $x^2$ direction
becomes two 4-branes positioned symmetrically with respect to $Z_2$ in
type-IA at specific $x^{11}$ coordinates.
These coordinates corresponds to the value of an $Sp(1)$ Wilson
loop around the $x^2$ direction in type-I.
We note that the $Sp(1)$ Wilson loop corresponds to a different
phase in the moduli space,
and the small instanton can be excited into this phase as well.
In 6D there are only the excitations by VEVs to the hyper-multiplets
in the fundamental representation $(\rep{2},\rep{32})$ which
describes the instanton (in the heterotic $SO(32)$) acquiring
a finite size \cite{WitSML}. Upon compactification to 5D,
the vector-multiplet, which had no scalars in 6D, has the Wilson
loop value as its scalar super-partner. When this Wilson loop is
turned on at a generic value, all the hyper-multiplets become
massive and thus have VEV zero. This means that we are on a different
phase.

Returning to type-IA, we find that an $Sp(1)$ Wilson loop (in type I)
splits the 4-brane (which was T-dual to the original 5-brane)
into two separate mirrored 4-branes.
The type IA came from an orbifold of type-\IIA,
 the latter being the \Mth\ wrapped on the 11th direction.
Thus we end up with two 5-branes in the \Mth\ at two different
coordinates in the $x^{11}$ direction (which is the direction
of the $\MS{1}/\IZ_2$). These two coordinates must be mirror
images of one another under the $\IZ_2$. We thus recovered
the \Mth\ picture, but we have also learned that separating
the two 5-branes in the \Mth\ (away from the orbifold fixed point
where the $E_8$ lives), corresponds not to an instanton of
finite size, but to a different phase. We elaborate on the phase
transition in section (5.2).

We can now substantiate the claim made before that the 
$(\rep{2},\rep{16})$ correspond to winding states of the
non-critical string. When a small $Sp(1)$ Wilson loop
of magnitude proportional to $\epsilon$ is turned on,
the hypermultiplets $(\rep{2},\rep{16})$ become massive
with a mass proportional to $\epsilon$.
On the other hand, we saw that
in the \Mth\ the 5-branes are now separated a distance proportional
to $\epsilon$ so the non-critical string connecting the
5-branes as in \cite{StrOPN} as well as the non-critical
string connecting the five-brane to the $\IZ_2$ fixed
point have now a tension 
proportional to $\epsilon$ in correspondence with the expected
mass of their winding states.

\subsection{$k$ instantons}
The above picture leads us to conclude that when an \ExE\ instanton
goes to zero size, a five brane emerges on one of the $Z_2$ fixed
points corresponding to one of the $E_8$ gauge groups in which the
instanton is embedded.
The analysis of the massless spectrum goes as follows.
We first consider the case in which the five-brane is separated from the
nine-brane. This corresponds, as we argued, to a $Sp(1)$ Wilson line after
compactification on a circle. In this case the massless spectrum contains one
tensor multiplet on the six dimensional world volume of the five brane.
The other case is when the five-brane is inside the nine-brane. This
corresponds to a small instanton.
We have no tensor multiplet but
instead a tensionless $5-9$ string emerges on the six dimensional
worldvolume from a membrane connecting the five-brane and the nine-brane.
In addition there is a $5-5$ non-critical tensionless string. This string
emerges from the intersection of a membrane which
connects the five brane with its own $Z_2$ image.
In both cases the string
tension is proportional to the distance between the branes.

This can be generalized to a configuration for which $k$ instantons
shrink to zero size at the same point. In the dual picture we have
four types of tensionless
strings. One which, as before, connects the 5-brane to the 9-brane.
There are $k$ such strings which transform in the fundamental
representation of $SO(k)$. The second type is a string which connects
two distinct 5-branes. There are $k(k-1)/2$ such strings and they
transform in the adjoint representation of $SO(k)$. The third type is
a string which connects a five-brane to a $Z_2$ image of another
distinct five brane. There are also $k(k-1)/2$ such strings and they
too transform in the adjoint representation of $SO(k)$.
The fourth type is a string which emerges from a membrane which connects a
five-brane with its image. There are $k$ such strings and they transform in
the fundamental representation of $SO(k)$.

In addition if we separate the five-branes from the nine-brane
(we can only give it a meaning after compactification on a circle and, as
discussed before, giving a Wilson line to the $Sp(k)$ group) the $5-9$
strings and the $5-5$ strings which connect two $Z_2$ images get nonzero
tension while $k$ massless tensor multiplets on the world volume of the
five-branes emerge. The tensors transform in the fundamental representation
of $SO(k)$.
There are still $k(k-1)/2$ tensionless $5-5$ strings
connecting the distinct five-branes.

When we consider the reduction of this configuration to type I by a
double zero radius limit, namely taking both $x^{2},x^{11}$ to be very
small, we should recover the dual picture for small instantons on the
$SO(32)$ heterotic string. We recall that when $k$ instantons shrink
to zero size in $SO(32)$ a nonperturbative $Sp(k)$ gauge group appears
with hypermultiplets in the antisymmetric representation of $Sp(k)$ and
also hypermultiplets in the $(\rep{2k},\rep{32})$
representation of $Sp(k)\times SO(32)$.
We see that if a $5-5$ ($5-9$) tensionless string gives rise to two (one) six
dimensional hypermultiplets
and two (one) six dimensional vector fields after compactification on 
$S^1\times S^1/Z_2$, then we reproduce the right spectrum for the gauge
group $Sp(k)$ and its antisymmetric representation.
The counting gives
$k+2\cdot [2\cdot k(k-1)/2+k]=k(2k+1)$ vector multiplets in the adjoint
of $Sp(k)$, and $k+2\cdot[2\cdot k(k-1)/2]=k(2k-1)$ in the
antisymmetric
representation of $Sp(k)$. The $k$ tensionless strings which connect
the 5 branes with the 9 brane will give rise to the remaining
hypermultiplets.

\subsection{Phase transitions via tensionless strings}
%% In six dimensions vector multiplets have no scalar components and
%% therefore there is no moduli space connected to vector multiplets.

In four dimensions,
different phases can be connected at points where
extra massless BPS particles appear \cite{SeiWit}.
The two phases differ by the number of
hypermultiplets and the number of vector multiplets.
In six dimensions there is no central charge for particles 
(this is related to the fact that vector multiplets have no scalar components)
and so such a mechanism can not bring about phase transitions.
However, there do exist central charges for 1-branes in 6D.
The BPS formula determines the tension of BPS 1-branes in terms
of expectation values of scalar components of tensor multiplets.
Can tensionless 1-branes connect two different phases?

In fact the tensionless string that we have discussed in the previous sections
is a transition point between two phases.
We can deform the small instanton in two ways. One way is to make the
instanton size nonzero. This deformation is parameterized by $29$
hypermultiplets as can verified by the dimension of the moduli space of an
$E_8$ instanton \cite{DMW}.
The other way is to separate the five-brane from the nine-brane thus giving the
non-critical string a tension. This phase has one extra tensor multiplet whose
scalar component corresponds to the $x^{11}$ distance of the five-brane to the
nine-brane. This phase was discovered in \cite{DMW} as a new way of cancelling
the anomaly equation. Passing from one phase to the other through the
tensionless string we trade $29$ hypers with $1$ tensor as the gravitational
anomaly in six dimensions asserts.

Upon compactification to $5$ dimensions the tensor multiplet becomes a vector
multiplet, tensionless strings become massless particles and the phase
transition mechanism becomes an analog of the four dimensional phase
transition.

\section{Discussion}
One of the by-products of the recent developments in nonperturbative
string theory is the discovery of a mysterious six-dimensional
low-energy theory which is not a field theory \cite{WitCOM}.
 This theory contains tensionless strings, but it is not clear
how to quantize them, neither is it clear whether
 the theory can be described by
quantizing tensionless strings.
There have been some suggestions for theories of tensionless
quantized strings \cite{TENSTR}.
\footnote{We are grateful to David Gross for bringing
these references to our knowledge.}
Recently, tensionless string theories have been constructed
directly from the \Mth\ \cite{StrOPN}, and indeed we have used
the ideas of the latter construction.

We have argued on the basis of T-duality that the small instanton
of the \ExE\ heterotic string might be described by a theory
which contains tensionless strings in 6D.
It seems that the tensionless strings arising from \ExE\
small instantons are novel in that they can couple to a six-dimensional
gauge field. (By taking $k$ 5-branes in
\cite{StrOPN} we obtain an $Sp(k)$ gauge group only {\em after
compactification} to 5D. The would-be gluons are 1-branes
in 6D.)

The 11 dimensional picture that we ``drew'' might be interpreted as a
bound-state of a 5-brane and a 9-brane. The mechanism for this binding
is related to tensionless string dynamics and it might be interesting
to explore it further.

It would be interesting to study the dynamics of the various low
energy excitations which emerge on the intersection of the five-branes and
nine-branes thus gaining more understanding on the strong coupling phase of
small \ExE\ instantons and the transition to the weak coupling phase.
It is plausible that the transition includes physics of tensionless strings.
One way of studying these strings is through a compactification to lower
dimensions.

One more point is in order.
When the instantons are embedded symmetrically in $E_8\times E_8$,
it was shown in \cite{DMW} that there exists a dual description
of the $\MR{6}$ physics in terms of a weakly coupled heterotic 
string. The non-perturbative $SU(2)$ which appears for a zero-size
instanton is mapped, in the dual description, to a perturbative
enhanced gauge symmetry that appears for special points in
the Narain moduli space.
 This picture appears to be very different than what we have
argued in the previous sections. In particular, we argued
that the 6D low-energy theory is a non-critical string
and this seems very remote from a field theory.
The reason for the discrepancy is that  in this paper we focussed
only on the contribution of the small instantons to the 6D low
energy spectrum. In order to obtain the {\em complete} 6D low energy
spectrum one needs to analyze the coupled system of the tensionless
strings with the rest of the fields (coming from the bulk of the K3 where
the interaction is through the $E_8$ gauge fields).
This problem involves tensionless string dynamics, and it is plausible
that as a result of that only the particle-like excitations of the
string remain.
It would be very interesting to understand under what circumstances
that happens.

We have seen a new type of phase transitions in six dimensions by the
appearance of a tensionless string.
In particular we have connected the phase of
a finite size instanton with the phase of a five-brane in the eleven dimensional
bulk in this way.
The appearance of a tensionless string seems to be important for understanding
other cases of phase transitions in six dimensions. In particular we recall
another situation in which a tensionless string seems to be relevant.
As was explained in \cite{DMW,WitNEW},
when the gauge bundle is embedded asymmetrically in \ExE, there is
a phase transition at strong coupling.
The description of this phase transition in terms of tensionless strings as
well as the role of tensionless strings in other phase transitions is under
current investigation \cite{US}.

\section*{Acknowledgements}
We wish to thank J. Blum,
P. Ho{\accent20 r}ava, K. Intriligator and
S. Ramgoolam for useful discussions.
We are very grateful to  David Gross for a very intriguing discussion
and especially grateful to E. Witten for explanations and help.
The research of OJG is supported by a Robert H. Dicke Fellowship.
The research of AH is supported by NSF Grant PHY92-45317.

\section*{Appendix: Gravitational anomalies on the 2-brane}
In section (4.2) we assumed that nothing special ``lives''
on the 2D intersection of the 5-brane and the 2-brane.
This was based on the fact that there are no gravitational
anomalies supported at that end.

Let's briefly recall how the anomalies were calculated
in \cite{HorWit}. They had a 2-brane stretched between
the two fixed points and they wanted to determine
what lives on the two 2D boundaries from a knowledge of the
behavior of the fields on the 2-brane bulk. The fields
on the 2-brane bulk are the reduction of $U(1)$ super YM in 10D
down to 3D (see \cite{WitBST}). In 3D we get 8 scalars (using
the fact that a vector is dual to a scalar in 3D) that
represent transverse oscillations together with their super-partners
that are spinors in the $(\rep{2},\rep{8}'')$ of $SO(1,2)\times SO(8)$
(where the notation of \cite{HorWit} was that $\rep{8},\rep{8}'$
and $\rep{8}''$ were the vector and two spinors of $SO(8)$ and $\rep{2}$
was the spinor of $SO(1,2)$. Those are the spinors that obey
$\dirac_1\dirac_{2}\dirac_{11} = 1$ which is the BPS condition of the
2-brane.
Since the boundary conditions that the bulk fields satisfy 
is known, the contribution of the bulk fields to the gravitational 
anomalies under a reparametrization
of the $x^1,x^{2},x^{11}$ coordinates can, in principle,
be determined.
In \cite{HorWit} a shortcut was made by arguing that the anomaly
is supported just at the ends so that it is sufficient to
consider reparametrization of the $x^1,x^{2}$ plane that is
independent of the $x^{11}$ coordinate, and the anomaly of the
latter is determined by the zero modes (in the $x^{11}$-direction)
 of the fields. It was also explained that the $\rep{8}''_{-}$
fermions produce the same gravitational anomalies as right-moving
RNS fermions and superconformal ghosts, because the $SO(8)$ degrees
of freedom are not really independent of the world-sheet, since
the $SO(8)$ is the structure group of the normal bundle of the 
world-sheet in 10D.

After this review of \cite{HorWit}, we wish to repeat their 
calculation for the case at hand of a 2-brane stretched between
a 5-brane and a $\IZ_2$ fixed point.  It is intuitively
clear that since the 5-brane has no $\IZ_2$ there is no 2D chirality
condition from that end, and we should not get any anomaly,
so that the total 2D anomaly should be half that of a 2-brane
stretched between two $\IZ_2$ fixed points. Let's see how
it happens.
At the fixed point, the boundary condition on the fermions reads:
\be
\dirac_{11}\psi(x^{11}=0) = \psi(x^{11}=0)
\ee
Since the bulk fermions satisfy the BPS condition
\be
\dirac_1\dirac_{2}\dirac_{11}\psi = \psi
\ee
(this determines what we mean by the $\rep{2}$ of $SO(1,2)$ above.)
On the 5-brane end there is no $\IZ_2$ projection condition,
but there is the condition
\be
\dirac_1\dirac_2\dirac_3\dirac_4\dirac_5\dirac_{6}
\psi(x^{11} = r) = \psi(x^{11} = r)
\ee
This  condition comes from the fact that the bosons of the
2-brane that describe oscillations in the directions transverse
to the 5-brane have Dirichlet boundary conditions at the 2D boundary
of the 2-brane. Supersymmetry relates these boundary conditions
to the fermions, but we have to remember that at the 2D boundary
only half the supersymmetry survives (compared with the 2-brane bulk)
because the 5-brane has its own BPS condition.
To find the massless fermions that a 2D observer sees we must
take the fields to be independent of $x^{11}$.
So, we end up with those fields in $(\rep{2},\rep{8}'')$
of $SO(1,2)\times SO(8)$ that satisfy the additional conditions
(which break $SO(1,2)\times SO(8)$):
\be
\dirac_{11}\psi =
\dirac_1\dirac_{2}\dirac_{11}\psi = 
\dirac_1\dirac_2\dirac_3\dirac_4\dirac_5\dirac_{6} \psi = \psi
\ee
as well as being in $\rep{8}''_-$.:
\be
\dirac_3\dirac_4\dirac_5\dirac_6\dirac_7\dirac_8\dirac_9\dirac_{10}
\psi = -\psi
\ee

Let's decompose those fermions under $SO(1,1)\times SO(4)\times SO(4)$.
The surviving spinors are
\be
(\rep{2},\rep{2}')_{-}
\label{surv}
\ee
The final step is as in the last paragraph
of chapter (2) of \cite{HorWit}.  Since the second $SO(4)$
(in the decomposition $SO(8)\supset SO(4)\times SO(4)$) decouples 
because it is in the directions transverse to the 5-brane
that is assumed fixed, we see that we need two times the anomaly
of fermions in the $\rep{2}_{-}$ where the $\rep{2}$ is the
negative chirality spinor of the normal bundle $N_4$ of the 2D
surface $\Sigma$ in the 5-brane bulk $M_6$ (i.e.
$TM_6 = N_4\oplus T\Sigma_2$ where subscripts denote dimensions).
The index for the gravitational
anomalies for a chiral spinor in the representation $\rep{r}$ is
\be
\hat{I}_\hlf (F,R)
=\trr{\rep{r}}{e^{iF}}(1+\inv{48}\tr{R^2}+\cdots).
\ee
In 2D the 4-form part of the above formula equals
\be
{{c-\bar{c}}\over {24}} \tr{R^2},
\ee
where $c$ and $\bar{c}$ are the left and right central charges 
respectively.
In our case $\rep{r} = \rep{2}$ and we also know that
in the fundamental representation of $SO(4)$ the bundle
$N\oplus T\Sigma$ is trivial so that
\be
\trr{\rep{4}}{F\wdg F} + \tr{R\wdg R} = 0.
\ee
Now we find
\bear
\trr{\rep{2}}{1} &=& 2, \nn\\
\trr{\rep{2}}{F\wdg F} &=& \inv{4}\trr{\rep{4}}{F\wdg F}
=-\inv{4}\tr{R\wdg R} .\nn
\eear
So we find 
\be
\hat{I}_\hlf (F,R)
=({2\over {48}} -\hlf\cdot (-\inv{4}))\tr{R^2}
=\inv{6}\tr{R^2}.
\ee
For comparison, the $\rep{1}$ representation would give
\be
\inv{48}\tr{R^2}.
\ee
So the contribution of $\rep{2}_{-}$ to the anomaly 
is like 8 left moving fermions i.e. $c-\bar{c} =4$.
Altogether the anomaly of \ref{surv} is twice as much, that is $8$.
(For a check let's see how the $\rep{8}''_{-}$ of \cite{HorWit}
gives $c-\bar{c} = 16$. We find
\bear
\trr{\rep{8}''}{1} &=& 8 \nn\\
\trr{\rep{8}''}{F\wdg F} &=& \trr{\rep{8}}{F\wdg F}
=-\tr{R\wdg R} \nn
\eear
\be
\hat{I}_\hlf (F,R)
=({8\over {48}} -\hlf\cdot (-1))\tr{R^2}
={{16}\over {24}}\tr{R^2}
\ee

\end{document}